\documentclass[prd,aps,showpacs,preprintnumbers,amsmath, amssymb]{revtex4-2}

\usepackage{graphics,epsfig,hyperref}
\usepackage{graphicx}
\usepackage{dcolumn}
\usepackage{bm}
\usepackage{epstopdf}
\usepackage{color}
\usepackage{xcolor}
\usepackage{subfigure}
\usepackage{diagbox} 
\usepackage{adjustbox}
\usepackage{multirow}
\usepackage{amsmath,amssymb,amsfonts}

\usepackage[normalem]{ulem}
\usepackage{xcolor}

\begin{document}
	\title{\boldmath Probing Kalb-Ramond field  with extreme mass ratio inspirals}
	
	\author{ Zhong-Wu Xia$^{1}$, Huajie Gong$^{1}$\footnote{huajiegong@hunnu.edu.cn}, Qiyuan Pan$^{1,2}$\footnote{panqiyuan@hunnu.edu.cn} and
		Jiliang Jing$^{1,2}$\footnote{jljing@hunnu.edu.cn}}
	\affiliation{$^1$Department of Physics, Institute of Interdisciplinary Studies, Key Laboratory of Low Dimensional Quantum Structures and Quantum Control of Ministry of Education, Synergetic Innovation Center for Quantum Effects and Applications, and Hunan Research Center of the Basic Discipline for Quantum Effects and Quantum Technologies, Hunan Normal University,  Changsha, Hunan 410081, People's Republic of China} 
	\affiliation{$^{2}$Center for Gravitation and Cosmology, College of Physical Science and Technology, Yangzhou University, Yangzhou 225009, People's Republic of China}

	\begin{abstract}
	\baselineskip=0.6 cm
	\begin{center}
		{\bf Abstract}
	\end{center}	
The extreme-mass-ratio inspirals (EMRIs) are emerging as precision laboratories for testing the gravity beyond general relativity. In this work, we investigate the Lorentz symmetry breaking (LSB) effect induced by the Kalb-Ramond (KR) field on the gravitational waveforms from the EMRI system. We observe that the  LSB parameter \( l \) appears in the leading order for the corrections of energy and angular momentum fluxes, and as \( |l| \) increases, the differences in EMRI waveforms between the KR black hole and Schwarzschild black hole become more pronounced. We note that the LSB effect becomes detectable by LISA for values of $|l|\sim 10^{-6}$ with a one-year observation period. Furthermore, we use the Fisher information matrix (FIM) approach for the parameter estimation and find the detection error for $l$ can be constrained to \(\Delta l \sim 10^{-5}\) at \(\mathrm{SNR} = 20\), demonstrating the potential of space-based gravitational wave detectors to rigorously test the KR field.

     \end{abstract}

\maketitle
\flushbottom

\section{Introduction}

The extreme mass-ratio inspiral (EMRI), the system where a stellar-mass compact object 
spirals into a supermassive black hole (SMBH) in the strong-field region, provides a rich source of information about the dynamics of strong-field gravity~\cite{cardenas2024testing}. This system is particularly valuable for exploring the extreme environments near SMBHs, as the small object acts as a precise probe, tracing the intricate spacetime geometry of the massive central black hole. 
The gravitational waves (GWs) emitted by EMRIs are expected to provide deep insights into not only the fine structure of the secondary compact object, including its spin~\cite{piovano2020extreme}, scalar charge~\cite{maselli2020detecting,guo2022detection}, vector charge~\cite{zhang2024probing}, and dipole moments~\cite{lestingi2024extreme}, but also the potential deviations from general relativity (GR), such as modifications to gravity~\cite{sopuerta2009extreme,pani2011gravitational} and interactions with dark matter~\cite{duque2024extreme} will be further explored by upcoming space-based detectors like the Laser Interferometer Space Antenna (LISA)~\cite{barausse2020prospects,Amaro-Seoane2023,colpi2024lisa}, TianQin~\cite{Luo2016,mei2021tianqin,luo2025progress}, and Taiji~\cite{hu2017taiji,Gong2019}. These detectors are designed to detect the millihertz GW signals emitted by EMRIs, which can last for months to years within the detector's sensitivity band, providing high signal-to-noise ratios (SNR) suitable for precision measurements.
Consequently, the EMRI observations will offer an unprecedented opportunity to test both GR and various modified gravity theories, potentially exceeding the constraints provided by current weak-field tests in Solar System~\cite{Li:2022pcy,Wagle:2023fwl,han2019testing,Rahman2024,yang2025gravitational,Cano:2024ezp,Tan:2024hzw,Gu:2024dna,Qiao:2024gfb,Fu:2024cfk,zhao2024dilated,Melis:2025iaw}.

On the other hand, the Lorentz symmetry, as a cornerstone of modern physics, may be violated in certain modified gravity theories, including the string theory~\cite{kostelecky1989spontaneous} and loop quantum gravity~\cite{alfaro2002loop}. The Kalb-Ramond (KR) field naturally emerges in the bosonic spectrum of string theory as a second-rank antisymmetric tensor field~\cite{kalb1974classical,yang2023static}. It has been extensively studied in various theoretical contexts, including black hole physics, cosmology, and braneworld scenarios. In effective field theories, the KR field serves as a mechanism for the spontaneous Lorentz symmetry breaking (LSB) when it acquires a nonzero vacuum expectation value (VEV). Specifically, the spontaneous LSB occurs when the KR tensor field is nonminimally coupled with the gravity through interactions involving the scalar curvature and the Ricci tensor. The nonzero VEV selects a preferred direction in spacetime, thereby spontaneously breaking the local Lorentz invariance of the vacuum state~\cite{altschul2010lorentz,bluhm2008spontaneous,liu202exact}. Physically, this spontaneous symmetry breaking typically arises in theoretical models that aim to describe physics near or beyond the Planck scale, where the classical Lorentz symmetry might no longer hold true~\cite{kostelecky1989spontaneous}. Furthermore, recent observations by the Event Horizon Telescope (EHT) of Sagittarius A*~\cite{event2023first,vagnozzi2023horizon} provide additional constraints on the Lorentz-violating parameter, restricting it to the range $-0.185022 \leq l \leq 0.060938$ \cite{Junior2024}. However, our current analysis using the EMRI demonstrates the capability to impose even tighter constraints on the KR parameter, i.e.,  $|l|\sim 10^{-6}$, compared to those derived from EHT observations. Therefore, the EMRI system offers a more precise probe of LSB effects. Investigating the KR field's impact on the gravitational lensing~\cite{Junior2024}, photon spheres~\cite{junior2024spontaneous}, shadows~\cite{Liu:2024lve,zahid2024shadow}, and quasinormal modes~\cite{Guo2024} can also provide a unique opportunity to probe LSB in strong gravitational fields.

Given that the EMRIs probe the strong-field regime of gravity with exceptional precision, they are particularly sensitive to the possible deviations introduced by the LSB. 
Ref. \cite{junior2024periodical} shows that the KR black hole spacetime significantly alters the dynamics of an EMRI system. 
Due to the deviations introduced by the LSB, the central black hole spacetime is no longer exactly described by the Schwarzschild metric. This modification influences the orbital motion of the secondary object around the central black hole, thereby affecting the energy and angular momentum radiated as gravitational waves.
Furthermore, Refs. \cite{Qiao:2024gfb,Fu:2024cfk,Zhang:2024csc,zi2025eccentric,Tan:2024utr,Guo:2025scs} suggest that the subtle deviations could potentially be detected through the high-precision analysis of EMRI gravitational waveforms. By using the modified flux formulas that include the effect of the KR field, it becomes possible to quantify the deviations from the expected Schwarzschild dynamics, offering a new method to constrain the parameters of KR field.

In this work, we investigate the effects of LSB in the KR black hole model on the EMRIs, where the LSB is characterized by a parameter $l$. Within this framework, we compute the energy and angular momentum fluxes of the system, and examine their influences on the orbital dynamics and gravitational waveforms. Our results indicate that the LSB parameter contributes at the leading order to the energy and angular momentum fluxes, and has significant impact on the orbital dynamics and gravitational waveforms. Furthermore, we assess the detectability of the LSB effects by EMRIs using two data analysis methods: the mismatch and the Fisher information matrix (FIM) approach.

This work is structured as follows. In Sec. \ref{section:KR_background}, we review the static KR black hole solution and analyze
the impact of LSB on the geodesic motion of the secondary object. In Sec. \ref{section:orbital_dynamics}, we delineate the detailed calculation of GW fluxes and orbital evolution. In Sec. \ref{section:GW_waveforms}, we presents a comprehensive
numerical analysis of the trajectory motion and the corresponding EMRI waveforms, and performs the data analysis.
Finally, we summarize and discuss our conclusions in the last section.

\section{Kalb–Ramond black hole and the motion of a particle}
\label{section:KR_background}
We begin by considering the KR field model, which is a representative example of LSB theories. The action of the model 
is given by~\cite{altschul2010lorentz,lessa2020modified,yang2023static}
\begin{equation}\label{action}
	S = \frac{1}{2\kappa} \int d^4x \sqrt{-g} \left[ R - 2\Lambda - \frac{1}{6} H_{\mu\nu\rho} H^{\mu\nu\rho} - V(B_{\mu\nu} B^{\mu\nu}) + \xi B^{\mu\nu} B_{\mu\nu} R \right],
\end{equation}
where $\kappa = 8 \pi G$ with the Newtonian constant of gravitation $G$. $R$ denotes the Ricci scalar, $\Lambda$ characterizes the cosmological constant,
$B_{\mu\nu}$ is a rank-2 antisymmetric tensor that describes the KR field~\cite{altschul2010lorentz,lessa2020modified,yang2023static}, and  $H_{\mu\nu\rho} = \partial_{[\mu} B_{\nu\rho]} $ represents the field strength. The potential $V(B_{\mu\nu} B^{\mu\nu})$ ensures a non-zero vacuum expectation value  for the KR field, thereby inducing LSB effects. The coupling constant $\xi$ controls the interaction between the KR field and the gravitational field. Assuming a static and spherically symmetric black hole solution under the influence of the KR field, from the action \eqref{action} we can obtain the spacetime metric \cite{yang2023static}
\begin{equation}\label{metric}
ds^2 = -f(r) dt^2 + g(r)^{-1} dr^2 + r^2 d\theta^2 + r^2 \sin^2 \theta \, d\varphi^2,
\end{equation}
with 
\begin{equation}
f(r) = g(r) = \frac{1}{1-l} - \frac{2M}{r},
\end{equation}
where $M$ is the mass of the black hole, and $l$ characterizes the LSB effect which has a range $-0.185022\leq l \leq 0.060938$ \cite{Junior2024}. For notational simplicity, we introduce the parameter $\lambda = l/(1-l)$.

For the motion of a test particle of mass \(m\) on the equatorial plane, we set $\theta=\pi/2$. Thus, we use the Lagrangian~\cite{chandrasekhar1998mathematical}
\begin{equation}
L = \frac{1}{2} m \left[-f(r) \dot{t}^2 + g(r)^{-1} \dot{r}^2 + r^2 \dot{\varphi}^2 \right],
\end{equation}
where the dot denotes differentiation with respect to the proper time \(\tau\). Given the symmetry of the spherically symmetric spacetime, there are two conserved quantities of geodesics: the energy \(E\) associated with time-translation invariance and the angular momentum \(L_z\) due to rotational symmetry, which can be derived from the Euler-Lagrange equations
\begin{equation}
	\begin{split}	
E = -\frac{\partial L}{\partial \dot{t}} &= m f(r) \dot{t},\\
L_z = \frac{\partial L}{\partial \dot{\varphi}} &= m r^2 \dot{\varphi}.
\end{split}
\label{eq:geodesic1}
\end{equation}
Substituting the above expressions (\ref{eq:geodesic1}) into the normalization condition for the four-velocity \(g_{\mu \nu} u^{\mu} u^{\nu} = -1\), we get the radial equation of motion
\begin{equation}
\dot{r}^2 = g(r) \left[ \frac{E^2}{m^2 f(r)} - \frac{L_z^2}{m^2 r^2} - 1 \right].\label{eq:geodesic2}
\end{equation}
For the eccentric motion, it is useful to introduce the  semi-latus rectum \(p\) and  eccentricity \(e\), and parametrize the radial coordinate \(r\) as~\cite{babak2007kludge}
\begin{equation}
	r(\chi) = \frac{p M}{1 + e \cos \chi},
\end{equation}
where \(\chi\) is the reparameterized radial coordinate, and \(p\) and \(e\) are the semi-latus rectum  and the eccentricity respectively.
At the orbit’s turning points (\(\dot{r} = 0\)), the radial coordinate $r(\chi)$ reaches its extrema: the periapsis $r_p$, corresponding to the minimum radius, and the apoapsis $r_a$, corresponding to the maximum radius, which are related to \(p\) and \(e\) via
\begin{equation}
r_p = \frac{pM}{1 + e}, \quad r_a = \frac{pM}{1 - e}.
\end{equation}
As a result, by solving \(\dot{r}(r_a) = 0=\dot{r}(r_p)\), the total energy \(E\) and the angular momentum \(L_z\) can be expressed as
\begin{equation}
	\begin{split}
E^2 &= m^2\frac{[(1+\lambda)p - 2]^2 - 4e^2}{p[(1+\lambda)p - 3 - e^2]},\\
L_z^2 &= \frac{m^2 M^2 p^2}{(1+\lambda)p - 3 - e^2}.
\end{split}
\end{equation}

The motion of the test particle is governed by the radial and angular frequencies \(\Omega_r\) and \(\Omega_\varphi\), which can be derived from the time intervals between successive pericenter and apocenter passages
\begin{equation}
		\Omega_r = \frac{2\pi}{T_r}, \quad \Omega_\varphi = \frac{\Delta \varphi}{T_r},
\end{equation}
with the radial period \(T_r\) and the change in the azimuthal angle $\varphi$ over a single radial period \(\Delta \varphi\)
\begin{equation}
	\begin{split}
T_r &= 2 \int_{r_{\text{min}}}^{r_{\text{max}}} dt=
 \int_{0}^{2\pi} \frac{dt}{dr}\frac{dr}{d\chi}d\chi,\\
\Delta \varphi &= 2 \int_{r_{\text{min}}}^{r_{\text{max}}} d\varphi=
 \int_{0}^{2\pi} \frac{d\varphi}{dt}\frac{dt}{dr}\frac{dr}{d\chi}d\chi.
\end{split}
\end{equation}
For the metric \eqref{metric} considered here, we have
\begin{align}	
	T_r &=\frac{2 M  \pi}{(1+\lambda)^{1/2}\left( 1 - e^2 \right)^{3/2}} p^{3/2}+\frac{6M  \pi}{(1 + \lambda)^{3/2} (1 - e^2)^{1/2}} p^{1/2}+ \mathcal{O}(p^{-1/2}),\\
	\Omega_r &= \frac{(1+\lambda)^{1/2}(1 - e^2)^{3/2}}{M} p^{-3/2} - \frac{3(1 - e^2)^{5/2}}{M(1+\lambda)^{1/2}} p^{-5/2} + \mathcal{O}(p^{-7/2}), \\
	\Omega_\varphi &= \frac{(1 - e^2)^{3/2}}{M} p^{-3/2} + \frac{3e^2 (1 - e^2)^{3/2}}{(1+\lambda)M} p^{-5/2} + \mathcal{O}(p^{-7/2}).
\end{align}

\section{GW fluxes and orbital evolution}
\label{section:orbital_dynamics}

In an EMRI system, a compact object gradually spirals into the SMBHs due to the emission of GWs. This GW emission carries away the energy and angular momentum from the system, leading to a gradual evolution of the orbit. We assume that the orbital evolution is adiabatic~\cite{isoyama2022adiabatic}. In this approximation, during a single orbital period, we ignore the backreaction caused by the radiation. At the end of each cycle, we take into account the cumulative changes occurred during that cycle and update the orbital parameters for the next cycle. 

The rates of change of the energy and angular momentum of the system due to the gravitational radiation are given by the energy flux $-\frac{dE}{dt}$ and angular momentum flux $-\frac{dL_i}{dt}$ respectively, which can be computed using the well-known quadrupole formula~\cite{thorne1980multipole,maggiore2008gravitational}
\begin{equation}
-\frac{dE}{dt} = \frac{1}{5} \left\langle \dddot{Q}_{ij} \dddot{Q}^{ij} \right\rangle,
\quad -\frac{dL_i}{dt} = \frac{2}{5} \epsilon_{ijk} \left\langle \ddot{Q}^{jm} \dddot{Q}^k_{\quad\!\!\!\! m} \right\rangle,\label{eq:flux}
\end{equation}
where the indices \(i, j\) correspond to the spatial components in Cartesian coordinates expressed as $(x^1,x^2,x^3)=(r\cos\varphi,r\sin\varphi,0)$. $Q_{ij} = m \left( x_i x_j - \frac{1}{3} \delta_{ij} x^kx_k \right)$ is the mass quadrupole moment of the system, \(\epsilon_{ijk}\) is the Levi-Civita symbol, and the angle brackets \(\langle \cdot \rangle\) represent averaging over one complete period. Substituting Eqs. (\ref{eq:geodesic1}) and (\ref{eq:geodesic2}) into Eq. (\ref{eq:flux}), we get the average energy and angular momentum fluxes for the Kalb–Ramond black hole
\begin{align}\label{eq:energy_flux}
\left\langle \frac{dE}{dt} \right\rangle =& \frac {(1 - e^2)^{3/2}m^2}{15p^5M^2}[96 + 292 e^2 + 37 e^4 - 3 e^2 \left(124 + 95 e^2 \right)\lambda \notag\\&+ 3 e^2 \left(20 + 73 e^2 + 10 e^4 \right)\lambda^2 + e^2 \left(4 + e^2 \right)\lambda^3
	]\notag\\
&+ \frac{e^2(1 - e^2)^{3/2}m^2}{5(1+\lambda)p^6M^2} [176 + 450 e^2 + 53 e^4 + \left(744 + 618 e^2 - 278 e^4 \right) \lambda \notag\\
&+ \left(-224 - 1314 e^2 - 335 e^4 + 30 e^6 \right) \lambda^2 
- 2 \left(12 + 21 e^2 + 2 e^4 \right) \lambda^3]
+ \mathcal{O}(p^{-7}),\\
\left\langle \frac{dL_z}{dt} \right\rangle =& \frac{4(1 - e^2)^{3/2}m^2}{5 p^{7/2}M}[8 + 7 e^2 - e^2 \left(14 + e^2 \right) \lambda + e^2 \left(3 + 2 e^2 \right) \lambda^2
]\notag\\
&+ \frac{4e^2(1 - e^2)^{3/2} m^2}{5 (1+\lambda)p^{9/2}M} [38 + 27 e^2 + \left(92 - 12 e^2 - 3 e^4 \right) \lambda \notag\\
&+ \left(-34 - 51 e^2 + 6 e^4 \right) \lambda^2
]
+ \mathcal{O}(p^{-11/2}).\label{eq:angular_flux}
\end{align}
It is easy to observe that the above expressions reduce to the Schwarzschild case when the parameter $\lambda=0$. Interestingly, the parameter $\lambda$ appears in the leading order of Eqs.~\eqref{eq:energy_flux} and~\eqref{eq:angular_flux}, indicating the dominating effect of the LSB parameter $l$.

\begin{figure}[tbp]
	\centering
	\begin{minipage}[b]{0.48\textwidth}
		\centering
		\includegraphics[width=\textwidth]{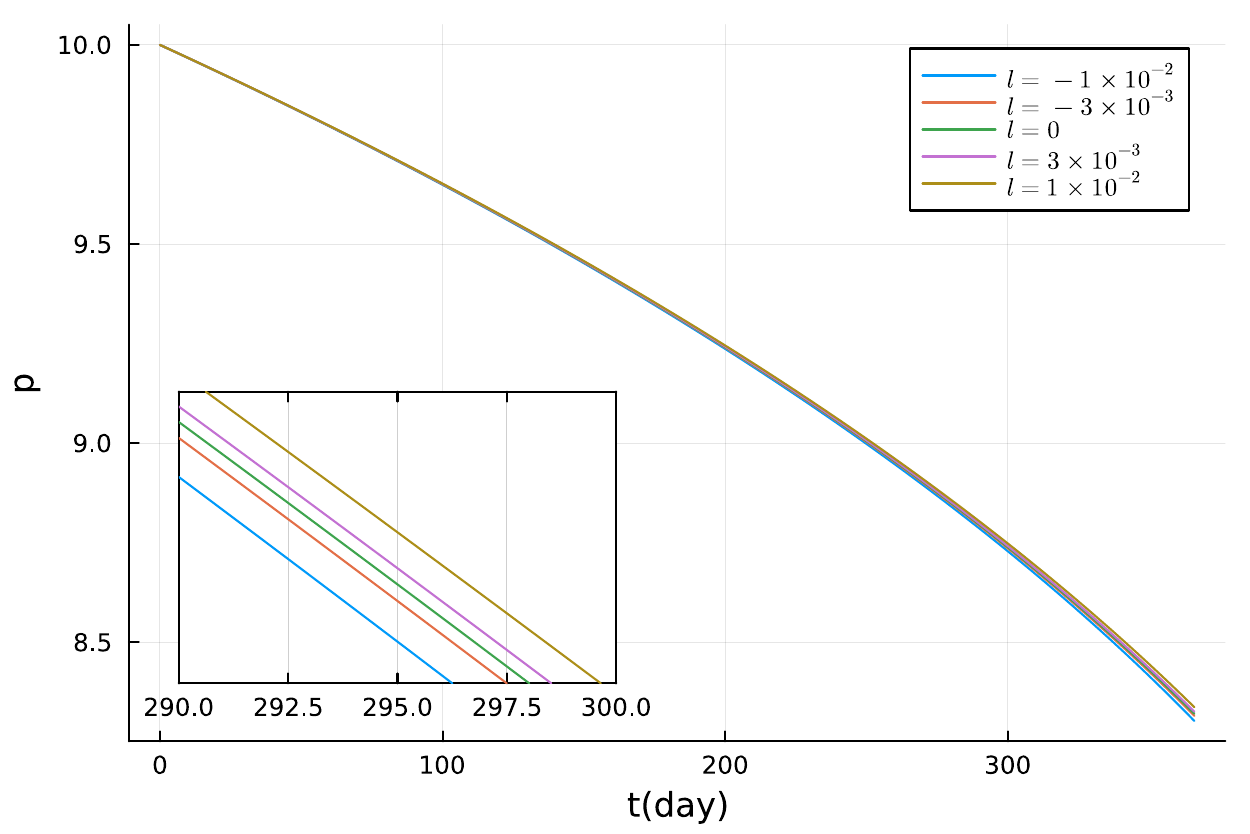}
	\end{minipage}
	\hfill
	\begin{minipage}[b]{0.48\textwidth}
		\centering
		\includegraphics[width=\textwidth]{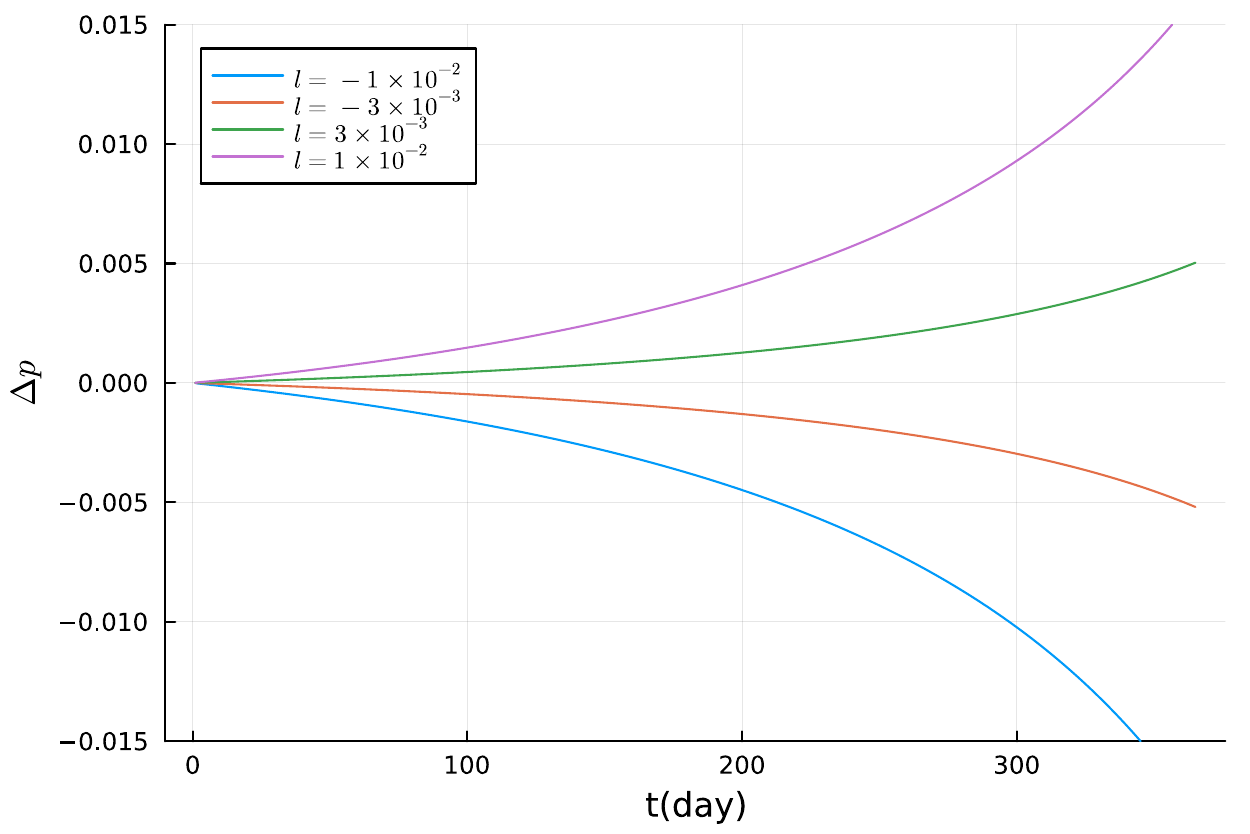}
	\end{minipage}
	\vspace{0.5cm} 
	
	\begin{minipage}[b]{0.48\textwidth}
		\centering
		\includegraphics[width=\textwidth]{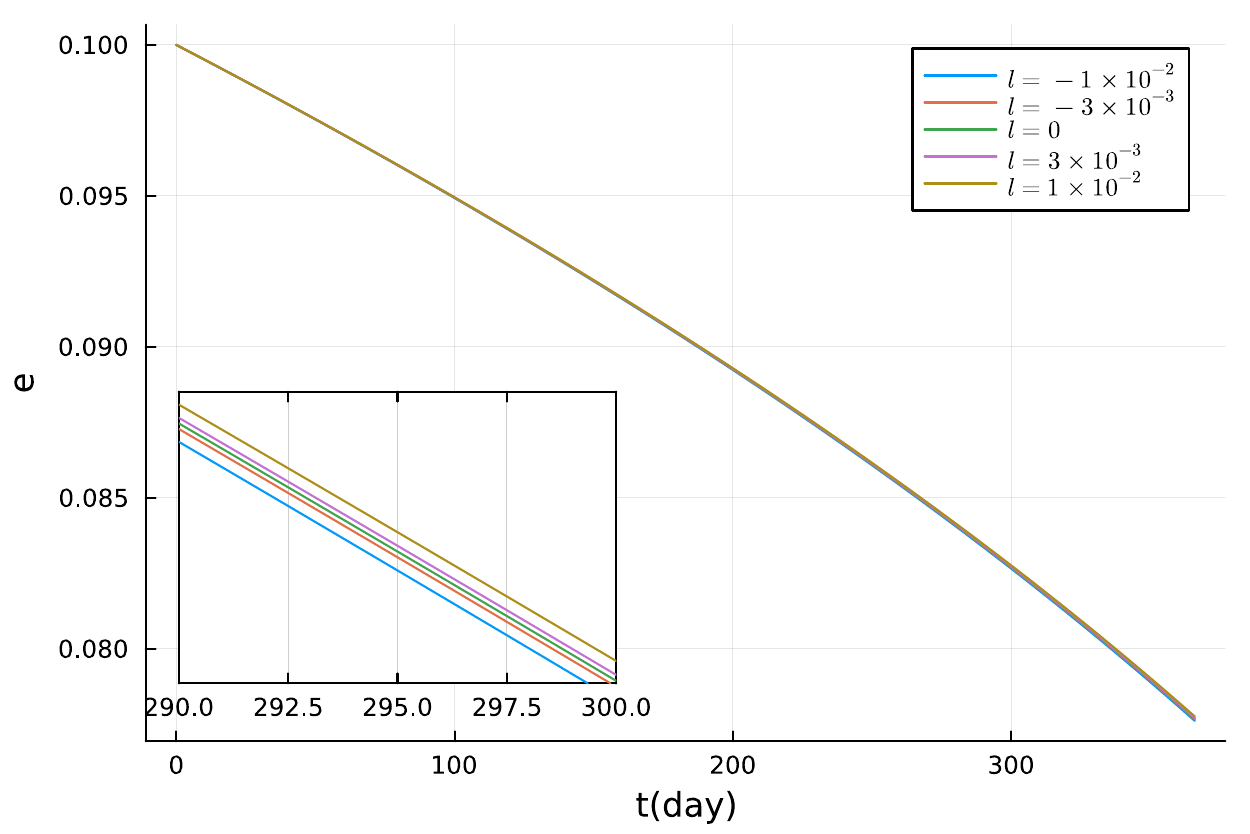}
	\end{minipage}
	\hfill
	\begin{minipage}[b]{0.48\textwidth}
		\centering
		\includegraphics[width=\textwidth]{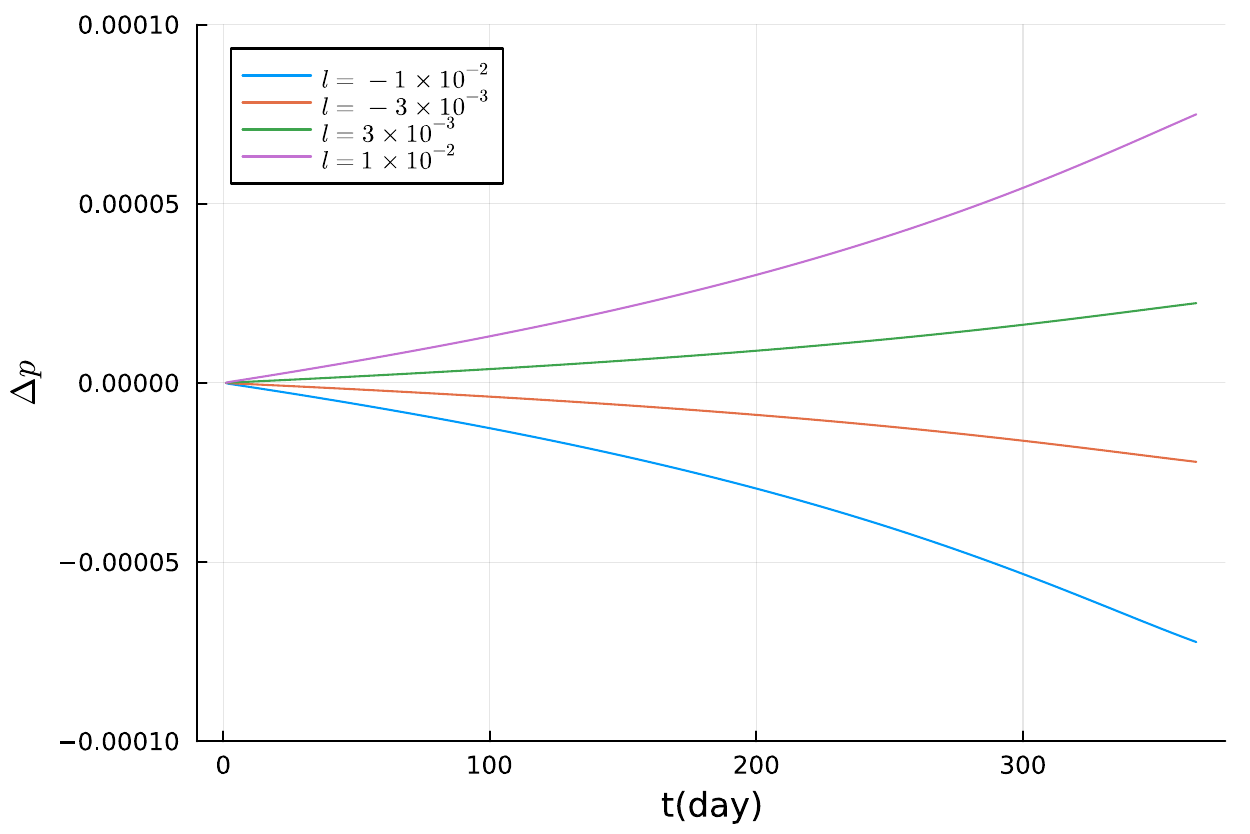}
	\end{minipage}
	
	\caption{The evolutions of the semi-latus rectum $p$ and eccentricity $e$ over time (left panels), and the differences $p^{(\text{KR})} - p^{(\text{Sch})}$ and $e^{(\text{KR})} - e^{(\text{Sch})}$ (right panels) for different LSB parameters $l$. Here, we set the initial values $(M,m,p,e)=(10^6M_\odot, 10M_\odot, 10M, 0.1)$. }
	\label{fig:orbital_evolution}
	
\end{figure}

The orbital evolution of the system can be obtained by integrating the energy and angular momentum fluxes over time. The total energy \(E\) and angular momentum \(L_z\) of the system depend on the semi-latus rectum \(p\) and eccentricity \(e\). Using the relationships
\begin{align}
\frac{dE}{dt} &= \frac{\partial E}{\partial p} \frac{dp}{dt} + \frac{\partial E}{\partial e} \frac{de}{dt},\\
\frac{dL_z}{dt} &= \frac{\partial L_z}{\partial p} \frac{dp}{dt} + \frac{\partial L_z}{\partial e} \frac{de}{dt},
\end{align}
we can derive the time evolution equations for \(p\) and \(e\)
\begin{align}
\left\langle \frac{dp}{dt}  \right\rangle=&
- \frac{8 \sqrt{1 + \lambda} \left(1 - e^2\right)^{3/2}  m}{5  p^3M}\left[8 + 7e^2- e^2 \left( 14 + e^2 \right) \lambda + e^2 \left( 3 + 2 e^2 \right) \lambda^2
\right]\notag\\
&+\frac{2 \left(1 - e^2\right)^{3/2} m}{15 \sqrt{1 + \lambda}  p^4M}\left[144 + 326 e^2 + 245 e^4 + 3 e^2 \left( 352 + 65 e^2 - 10 e^4 \right) \lambda \right.\notag\\
&\left.+ \left( -378 e^2 - 789 e^4 + 30 e^6 \right) \lambda^2 
- e^2 \left( 4 + e^2 \right) \lambda^3
\right]+ \mathcal{O}(p^{-5}),\label{eq:dpdt}\\
\left\langle \frac{de}{dt}  \right\rangle  =& -\frac{\sqrt{1+\lambda}e(1-e^2)^{3/2}m}{15p^4M}
\left[304+121e^2- 3 \left( 68 + 147 e^2 + 4 e^4 \right) \lambda \right.\notag\\
&
\left.+ 3 \left( 8 + 77 e^2 + 18 e^4 \right) \lambda^2 
+ \left( 4 + e^2 \right) \lambda^3\right]
-\frac{e(1-e^2)^{3/2}m}{30\sqrt{1+\lambda}p^5M}\notag\left[1280 + 2097 e^2 + 697 e^4 \right.\\ &+\left( 1236 + 5691 e^2 - 363 e^4 - 60 e^6 \right) \lambda 
+ \left( -408 - 6585 e^2 - 4215 e^4 + 150 e^6 \right) \lambda^2 \notag
\\&+ \left.\left( -124 - 267 e^2 - 29 e^4 \right) \lambda^3
\right]+ \mathcal{O}(p^{-6})\label{eq:dedt},
\end{align}
where the parameter $\lambda$ appears directly in the leading order, indicating a significant impact of the LSB effect on the orbital evolution. In Fig. \ref{fig:orbital_evolution}, we give the time evolution of an orbit with an initial semi-latus rectum $p=10M$ and eccentricity $e=0.1$ with $(M,m)=(10^6M_\odot, 10M_\odot)$ for different LSB parameters $l$ (corresponding to the parameters $\lambda = l/(1-l)$), i.e., $l=0, \pm 3\times 10^{-3}$ and $\pm 1\times 10^{-2}$. 
The left two panels show that, after one year, the semi-latus rectum decreases to approximately $p \sim 8.3M$, and the eccentricity reduces to $e \sim 0.078$. This behavior aligns with the classical prediction that the GW emission causes the orbital radius to shrink and the orbit to become more circular. The right two panels depict the evolution of the differences in orbital parameters between the KR black hole and the Schwarzschild black hole over time. It is evident that as the time $t$ progresses or the absolute value of the LSB parameter $l$ increases, the discrepancies in the orbital parameters also grow, which lead to the differences of inspiral phase between the KR black hole and the Schwarzschild black hole. Obviously, Eqs.~\eqref{eq:dpdt} and~\eqref{eq:dedt} describe the adiabatic inspiral of the compact object under the influence of gravitational radiation.

\section{Waveform and data Analysis}
\label{section:GW_waveforms}

To obtain accurate waveforms for EMRIs, a variety of waveform models have been developed, such as the analytic kludge (AK) \cite{peters1963gravitational,peters1964gravitational}, the numerical kludge (NK)~\cite{babak2007kludge}, and the adiabatic waveforms~\cite{hughes2021adiabatic,isoyama2022adiabatic,nasipak2024adiabatic}, with extensive studies devoted to improving their fidelity and computational efficiency~\cite{khalvati2025impact}. Among these, the augmented analytic kludge (AAK) model \cite{chua2017augmented} stands out by enhancing the traditional AK approach  through the incorporation of frequency corrections derived from the NK model, thereby ensuring high-fidelity waveforms while maintaining computational efficiency. This balance between efficiency and accuracy makes it an ideal choice for the upcoming LISA data analysis challenges, where large numbers of waveform templates need to be evaluated. Although the fully relativistic self-force waveforms offer the highest precision for modeling EMRIs~\cite{katz2021fast,chapman2025fast}, the limitations in current methods for modified gravity scenarios~\cite{Wagle:2023fwl} necessitate adopting the post-Newtonian (PN) expansions. Nevertheless, these expansions remain robust~\cite{babak2007kludge}, ensuring the reliable characterization of EMRI systems near the strong-field regime. Following this, we outline the statistical techniques used for parameter estimation, including methods such as the maximum likelihood estimation, which are applied to derive accurate astrophysical parameters.

\subsection{EMRI waveform}

\begin{figure}[tbp]
	\centering 
	\begin{minipage}[b]{1\textwidth}
		\centering
		\includegraphics[width=1\textwidth]{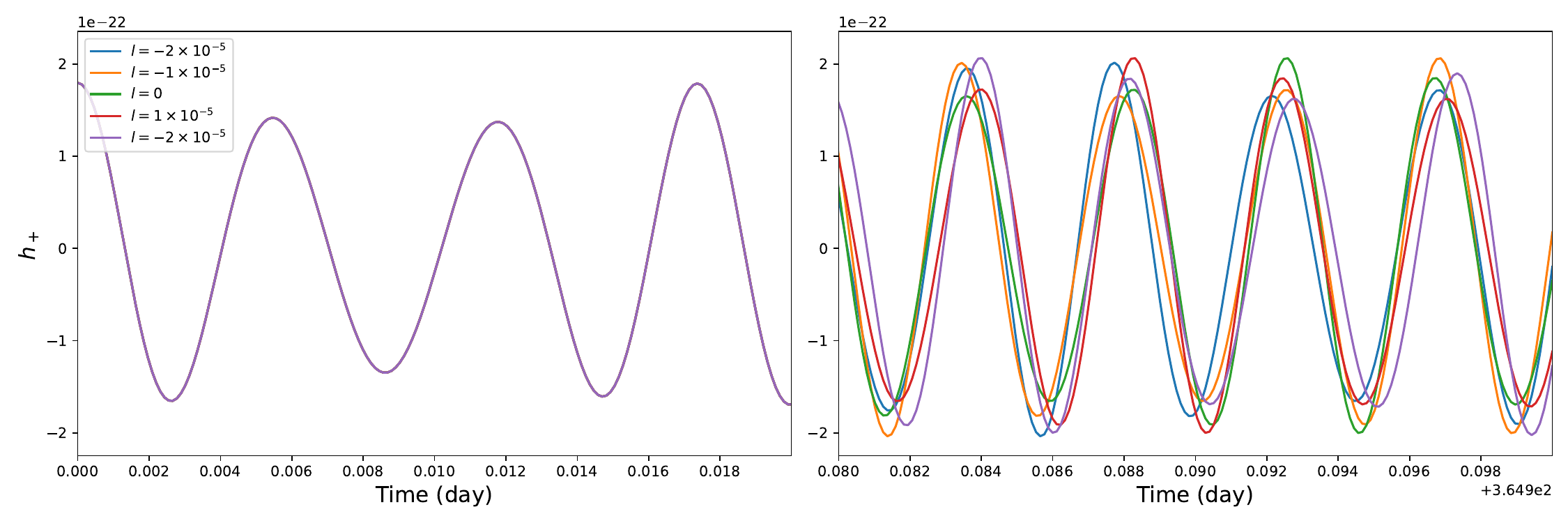}
	\end{minipage}
	\begin{minipage}[b]{1\textwidth}
		\centering
		\includegraphics[width=1\textwidth]{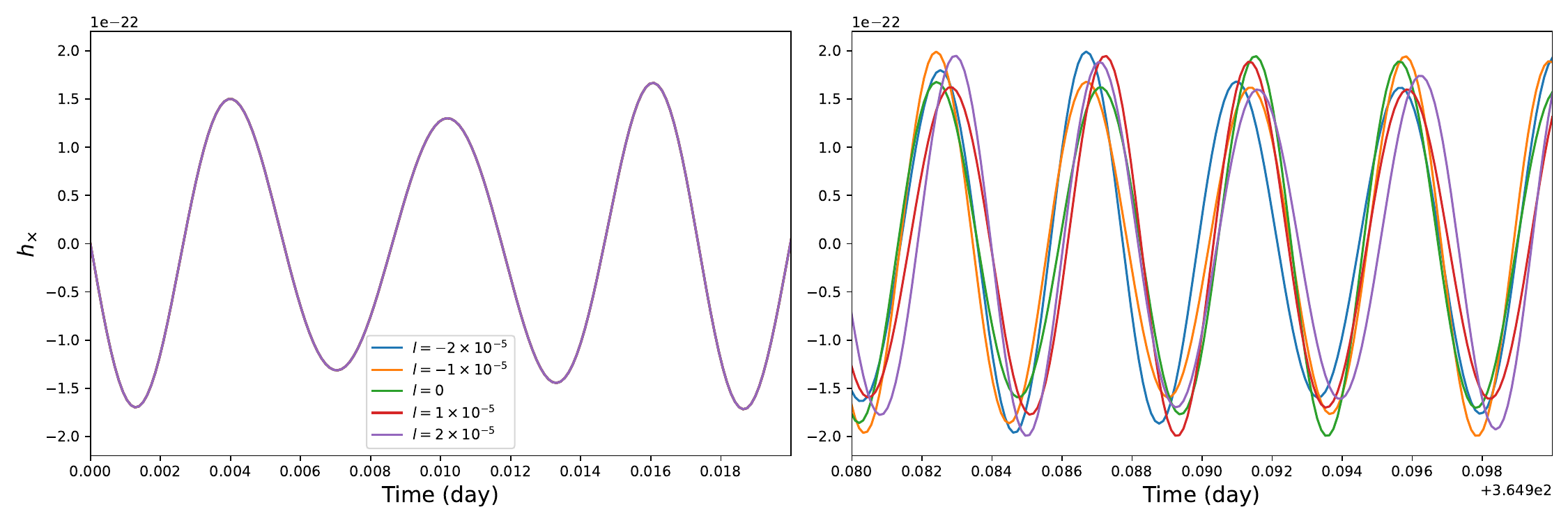}
	\end{minipage}
	\caption{\label{fig:waveform}  The EMRI waveforms for different LSB parameters $l$. Here, we set the initial values $(M,m,p,e)=(10^6M_\odot, 10M_\odot, 10M, 0.1)$. 
	}
\end{figure}
Once the inspiral geodesic orbits are obtained, the quadrupole approximation formula for the EMRI waveform can be expressed  in the transverse-traceless (TT) gauge as follows
\begin{equation}
	h_{ij} = \frac{2}{D} \left( P_{ik} P_{jl} - \frac{1}{2} P_{ij} P_{kl} \right) \ddot{I}^{kl},
\end{equation}
where \( D \) is the distance to the source, \( P_{ij} \) is the projection operator defined as \( P_{ij} = \delta_{ij} - \hat{n}_i \hat{n}_j \), and \( \ddot{I}^{kl} \) is the second time derivative of the mass quadrupole moment tensor with the coordinate time. For a binary system in a Keplerian orbit, the quadrupole moment tensor \( I^{ij} \) can be decomposed into harmonic components of the orbital frequency.  The gravitational wave signal can be described by two polarizations, and two polarization modes \( h_+ \) and \( h_\times \) can be expressed as a sum over all harmonics
\begin{equation}
	h_{+,\times} = \sum_{n=1}^\infty A_n^{+,\times}
\end{equation}
with
\begin{align}
	A_n^+ &= \left[ 1 + (\hat{L} \cdot \hat{n})^2 \right] \left[ b_n \sin(2\gamma) - a_n \cos(2\gamma) \right] + \left[ 1 - (\hat{L} \cdot \hat{n})^2 \right] c_n, \label{eq:hplus}\\
	A_n^\times &= 2 (\hat{L} \cdot \hat{n}) \left[ b_n \cos(2\gamma) + a_n \sin(2\gamma) \right],\label{eq:hcross}
\end{align}
where \( \gamma = \Phi_\phi(t) - \Phi_r(t) \) with the phase in the radial direction $\Phi_r(t)=\int_{0}^{t}\Omega_r(\xi)d\xi$ and the phase in the azimuthal direction $\Phi_\phi(t)=\int_{0}^{t}\Omega_\phi(\xi)d\xi$, and $
\hat{L} \cdot \hat{n} = \cos(\theta_K) \cos(\theta_S) + 
\sin(\theta_K) \sin(\theta_S) \cos(\phi_K- \phi_S)$ with the spherical polar angles $(\theta^{}_{\rm S},\phi^{}_{\rm S})$ describing the EMRI’s sky position in the Solar System barycenter frame while the angles $(\theta^{}_{\rm K},\phi^{}_{\rm K})$ determining the orientation of the SMBH spin in the same frame~\cite{pani2011gravitational}. In the following numerical calculations, we set the parameters $\theta_S=\pi/3$, $\phi_S=\pi/2$, $\theta_K=\pi/4$ and $\phi_K=\pi/4$. It is pertinent to mention that $T=1$yr represents the period of the Earth's orbit around the Sun. The coefficients \( a_n \), \( b_n \) and \( c_n \) are given by
\begin{align}
	a_n &= -n \mathcal A \left[ J_{n-2}(ne) - 2e J_{n-1}(ne) + \frac{2}{n} J_n(ne) + 2e J_{n+1}(ne) - J_{n+2}(ne) \right] \cos(n\Phi_r(t)), \\
	b_n &= -n \mathcal A \sqrt{1 - e^2} \left[ J_{n-2}(ne) - 2J_n(ne) + J_{n+2}(ne) \right] \sin(n\Phi_r(t)), \\
	c_n &= 2 \mathcal A J_n(ne) \cos(n\Phi_r(t)),
\end{align}
with the Bessel functions of the first kind \( J_n(ne) \). We have defined $\mathcal A=\left(M\Omega_\phi(t)\right)^{2/3}m/D$ with the luminosity distance $D=1\mathrm{Gpc}$. 

In our numerical calculations, we set the primary mass as $M=1\times 10^6M_\odot$ and the secondary mass as $m=10M_\odot$, resulting in a mass ratio of $q=m/M=10^{-5}$. Our choices of initial parameters $(p, e) = (10M, 0.1)$ take into account both the physical relevance and numerical reliability: on the one hand, these choices are consistent with recent astrophysical predictions for observable EMRIs~\cite{pan2021wet}; on the other hand, they ensure the computational validity and accuracy of the AAK waveform model~\cite{chua2017augmented}. The  difference in the LSB parameter directly leads to differences in the waveforms. As shown in Fig. \ref{fig:waveform}, at \( t = 0 \), the waveforms of the KR black hole and the Schwarzschild black hole almost coincide and are indistinguishable. However, due to the cumulative effects of radiation reaction and dephasing, their waveforms separate significantly after one year. Moreover, as the absolute value of LSB parameter \( l \) increases, the deviation relative to the Schwarzschild black hole becomes greater.

\subsection{LISA data analysis}

In practical EMRI data analysis, the detector's response to an incident GW can be expressed as
\begin{eqnarray}
	h_{\alpha}(t)=\frac{\sqrt{3}}{2}\left[F^{+}_{\alpha}(t)h^{}_{+}(t)+F^{\times}_{\alpha}(t)h^{}_{\times}(t)\right]\,,
\end{eqnarray}
with an index $\alpha$ denoting the distinct, independent channels of the detector. 
For the LISA detector, two independent Michelson-like interferometer channels are formed from the data stream. This dual-channel configuration allows for the capture of two distinct gravitational waveforms, thereby improving the reliability of our conclusions. As outlined in~\cite{barack2004lisa}, the antenna pattern functions, which characterize the detector's responses, are given by
\begin{align}
	F^{+}_I &= \frac{1}{2}(1+\cos^2\Theta)\cos(2\Phi)\cos(2\psi) -\cos\Theta\sin(2\Phi)\sin(2\psi)\,,  \nonumber \\
	F^{\times}_I &= \frac{1}{2}(1+\cos^2\Theta)\cos(2\Phi)\sin(2\psi)+\cos\Theta\sin(2\Phi)\cos(2\psi)\,, \nonumber \\
	F^{+}_{II} &= \frac{1}{2}(1+\cos^2\Theta)\sin(2\Phi)\cos(2\psi) +\cos\Theta\cos(2\Phi)\sin(2\psi)\,, \nonumber \\
	F^{\times}_{II} &= \frac{1}{2}(1+\cos^2\Theta)\sin(2\Phi)\sin(2\psi) -\cos\Theta\cos(2\Phi)\cos(2\psi)\,,
\end{align}
with the angles $(\Theta(t),\Phi(t))$ ~\cite{cutler1998angular}
\begin{align}
	\cos\Theta^{}(t) &= \frac{1}{2}\cos\theta^{}_{\rm S}-\frac{\sqrt{3}}{2}\sin\theta^{}_{\rm S}\cos\left(\frac{2\pi t}{T}-\phi^{}_{\rm S}\right) \,,\nonumber \\
	\Phi^{}(t)& = \frac{2\pi t}{T}+\tan^{-1}\left[\frac{\sqrt{3}\cos\theta^{}_{\rm S}+\sin\theta^{}_{\rm S}\cos\left(2\pi t/T -\phi^{}_{\rm S}\right)}{2\sin\theta^{}_{\rm S}\sin\left(2\pi t/T -\phi^{}_{\rm S}\right)}\right]\,,
\end{align}
and the polarization angle $\psi$
\begin{align}
	\tan \psi =& \left[ \left\{ \cos\theta^{}_{\rm K} -
	\sqrt{3}\sin\theta^{}_{\rm K}\cos\left(2\pi t/T -\phi^{}_{\rm K}\right)\right\}-2\cos\Theta(t)\left\{\cos\theta^{}_{\rm K}\cos\theta^{}_{\rm S} \right.\right. \nonumber\\ 
	&\left.\left.+\sin\theta^{}_{\rm K}\sin\theta^{}_{\rm S}\cos(\phi^{}_{\rm K}-\phi^{}_{\rm S}) \right\} \frac{}{}\right]/\left[\frac{}{} \sin\theta^{}_{\rm K}\sin\theta^{}_{\rm S}\sin(\phi^{}_{\rm K}-\phi^{}_{\rm S})\right.\nonumber\\ 
	&\left. -\sqrt{3}\cos\left(2\pi t/T \right)\left\{\cos\theta^{}_{\rm K}\sin\theta^{}_{\rm S}\sin\phi^{}_{\rm S}-\cos\theta^{}_{\rm S}\sin\theta^{}_{\rm K}\sin\phi^{}_{\rm K} \right\} \right.\nonumber\\
	&\left. -\sqrt{3}\sin\left(2\pi t/T\right) \left\{\cos\theta^{}_{\rm S}\sin\theta^{}_{\rm K}\cos\phi^{}_{\rm K}-\cos\theta^{}_{\rm K}\sin\theta^{}_{\rm S}\cos\phi^{}_{\rm S} \right\} \frac{}{}\right].\label{LISApsi}
\end{align}

To distinguish the difference of the waveforms between the Schwarzschild spacetime and KR
spacetime, we calculate the mismatch.
The mismatch function quantifies the dissimilarity between a signal waveform \( h_S \) and a template waveform \( h_T \) used in the matched filtering.  A small mismatch between the template and signal waveforms ensures that the template closely matches the true waveform, allowing for the accurate parameter extraction via the matched filtering. The mismatch is defined as
\begin{equation}
	\mathcal{M}(h_T, h_S) = 1-\frac{\langle h_T | h_S \rangle}{\sqrt{\langle h_T | h_T \rangle \langle h_S | h_S \rangle}},
\end{equation}
where the inner product \( \langle h_T | h_S \rangle \) is defined in the frequency domain as~\cite{cutler1998angular}
\begin{equation}
	\langle h_T | h_S \rangle = 4 \, \text{Re} \int_0^\infty \frac{\tilde{h}_T(f) \tilde{h}_S^*(f)}{S_n(f)} df.
\end{equation}
Here, \( \tilde{h}(f) \) represents the Fourier transform of the respective waveforms, and \( S_n(f) \) denotes the one-sided noise power spectral density of the detector~\cite{robson2019construction}. 
The mismatch function reaches its minimum value of zero when the template matches the signal perfectly in the presence of noise. Thus, minimizing the mismatch is a key step in optimizing template banks used in gravitational wave searches.

\begin{figure}[tbp]
	\centering 
	\begin{minipage}[b]{.49\textwidth}
		\centering
		\includegraphics[width=1\textwidth]{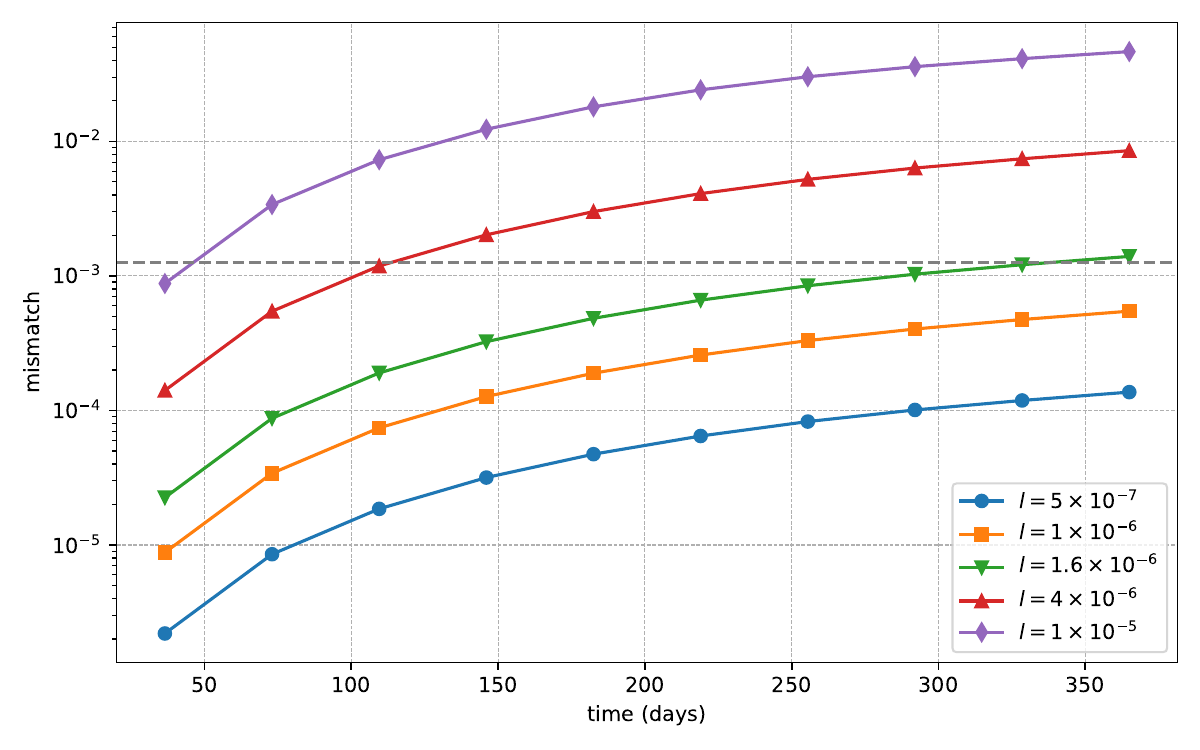}
	\end{minipage}
		\begin{minipage}[b]{.49\textwidth}
		\centering
		\includegraphics[width=1\textwidth]{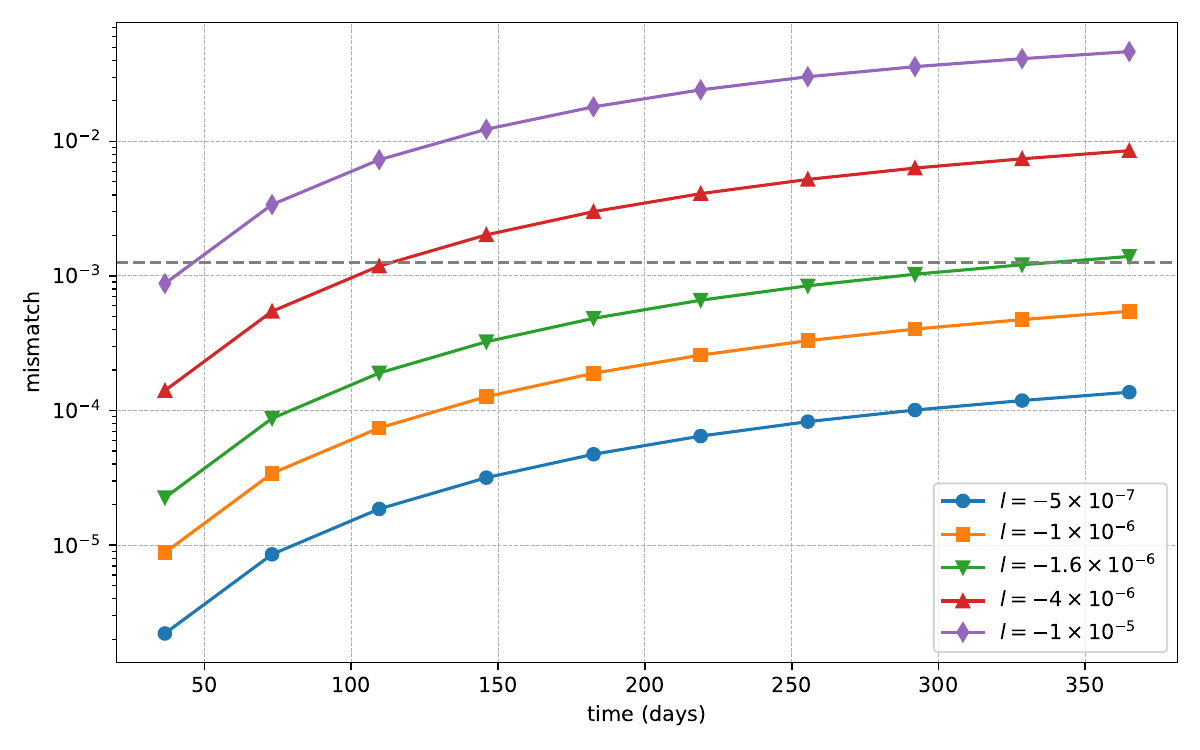}
	\end{minipage}
	\caption{\label{fig:mismatch-t} The mismatch as a function of observation time for different values of the LSB parameter \( l \). The horizontal black dashed line denotes the detection threshold of mismatch $\mathcal M=0.00125$.  Here, we set the initial values $(M,m,p,e)=(10^6M_\odot, 10M_\odot, 10M, 0.1)$.}
\end{figure}

\begin{figure}[tbp]
	\centering 
	\begin{minipage}[b]{.49\textwidth}
		\centering
		\includegraphics[width=1\textwidth]{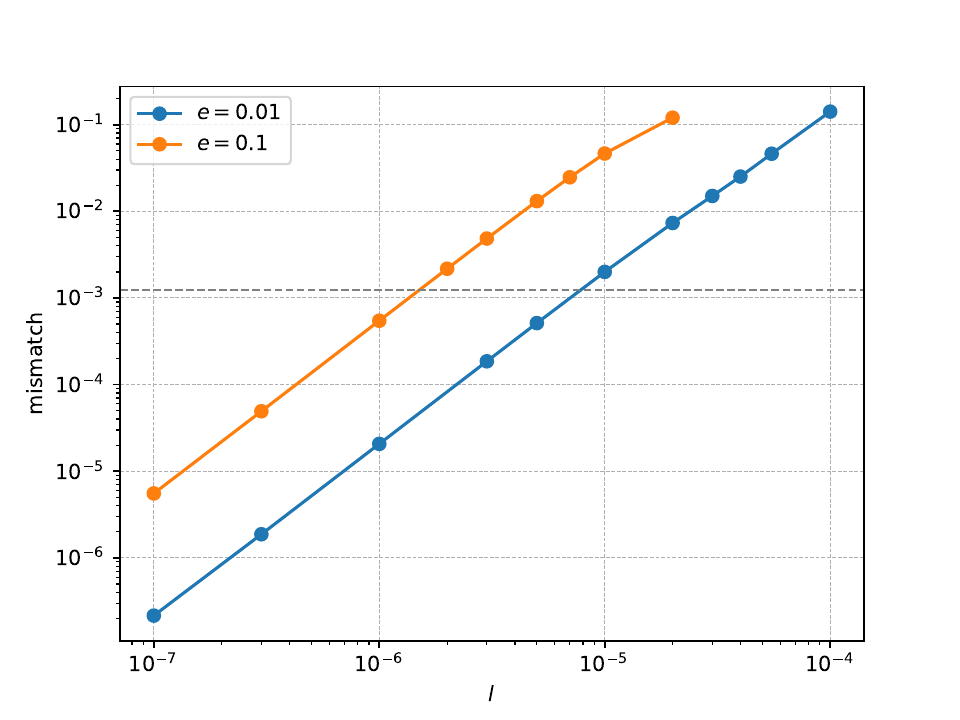}
	\end{minipage}
		\begin{minipage}[b]{.49\textwidth}
		\centering
		\includegraphics[width=1\textwidth]{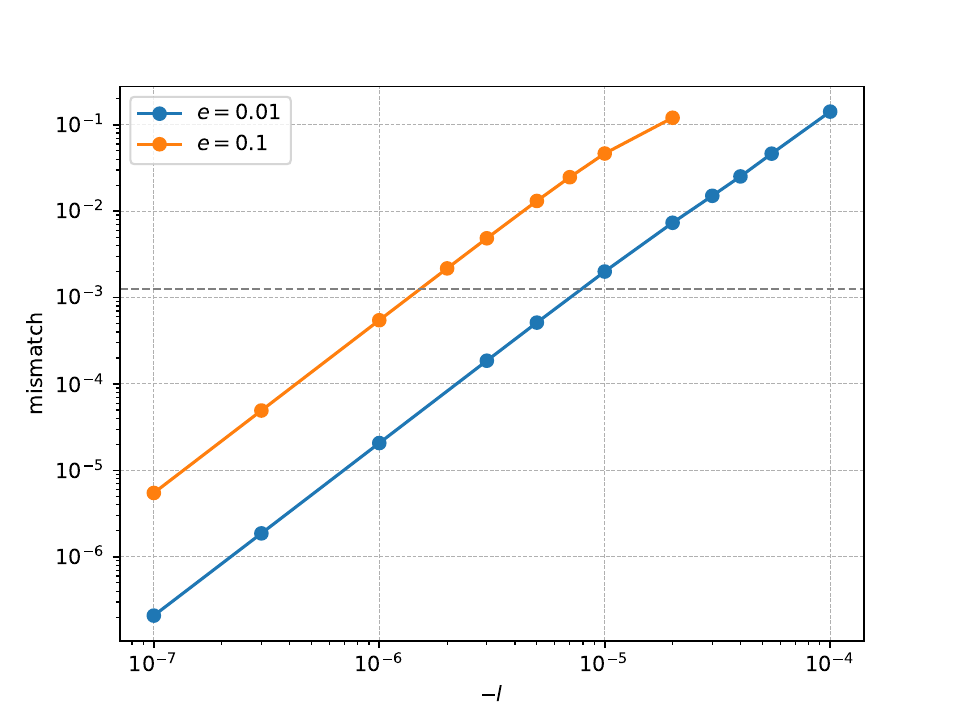}
	\end{minipage}
	\caption{\label{fig:mismatch-l} The one-year mismatch as a function of the LSB parameter \( l \) for two different values of the eccentricity \( e = 0.01 \) (blue) and \( e = 0.1 \) (orange). The horizontal black dashed line represents the detection threshold of mismatch $\mathcal M=0.00125$. Here, we set the initial values $(M,m,p,e)=(10^6M_\odot, 10M_\odot, 10M, 0.1)$.}
\end{figure}


To quantify the evolution of the waveform deviation over time,  we illustrate the mismatch between KR black hole and Schwarzschild black hole as a function of time in Fig. \ref{fig:mismatch-t}.
The mismatch is computed by observing its evolution over one-year period.
As shown in the figure, the mismatch increases with the observation time, and the growth rate of the mismatch becomes more pronounced as the absolute value of \( l \) (i.e., \( |l| \)) increases. For smaller \( |l| \) (e.g., \(5 \times 10^{-7}\)), the mismatch remains below the detection threshold for one-year period. However, as \( |l| \) increases to \(1.6 \times 10^{-6}\), the mismatch slightly surpasses the detection threshold $\mathcal M=0.00125$. For larger values of \( |l| \) (e.g., \( 1 \times 10^{-5} \)), the mismatch rapidly exceeds the threshold, indicating that the larger value of \( |l| \) leads to the more obvious deviation in the EMRI waveform. This behavior underscores the importance of the LSB parameter in distinguishing EMRI waveforms with LISA's observational constraints.

In addition to the LSB parameter, the eccentricity of the orbit also plays a crucial role in the evolution of the mismatch.
As shown in the Fig.~\ref{fig:mismatch-l}, the mismatch increases with the increase of the eccentricity \( e \). For a fixed value of \( l \), the mismatch is clearly larger for the higher eccentricity. For instance, if \( l = 3 \times 10^{-6} \), the mismatch is \( \mathcal{M} = 1.86 \times 10^{-4} \) for $e = 0.01$, whereas  \( \mathcal{M} = 4.85 \times 10^{-3} \) for $e = 0.1$. This shows that a larger eccentricity leads to a more pronounced mismatch for the same \( l \). From Fig.~\ref{fig:mismatch-l}, we observe that for the smaller eccentricity (e.g., \( e = 0.01 \), blue curve), the mismatch remains below the detection threshold for the entire one-year observation period even for the larger \( |l| \), e.g., \( l = 7 \times 10^{-6} \). However, as \( e \) increases (e.g., \( e = 0.1 \), orange curve), the mismatch surpasses the detection threshold even for the smaller \( |l| \), e.g., \( l = 3 \times 10^{-6} \). This suggests that, even for the smaller \( |l| \), the higher eccentricity leads to the larger deviation in the EMRI waveform. Thus, the combination of the LSB effect and the eccentricity provides richer physics in the EMRI waveforms, and the higher eccentricity makes it more easy for the detection of LSB effect.


\begin{figure}[tbp]
	\centering 
	\begin{minipage}[b]{.7\textwidth}
		\centering
		\includegraphics[width=1\textwidth]{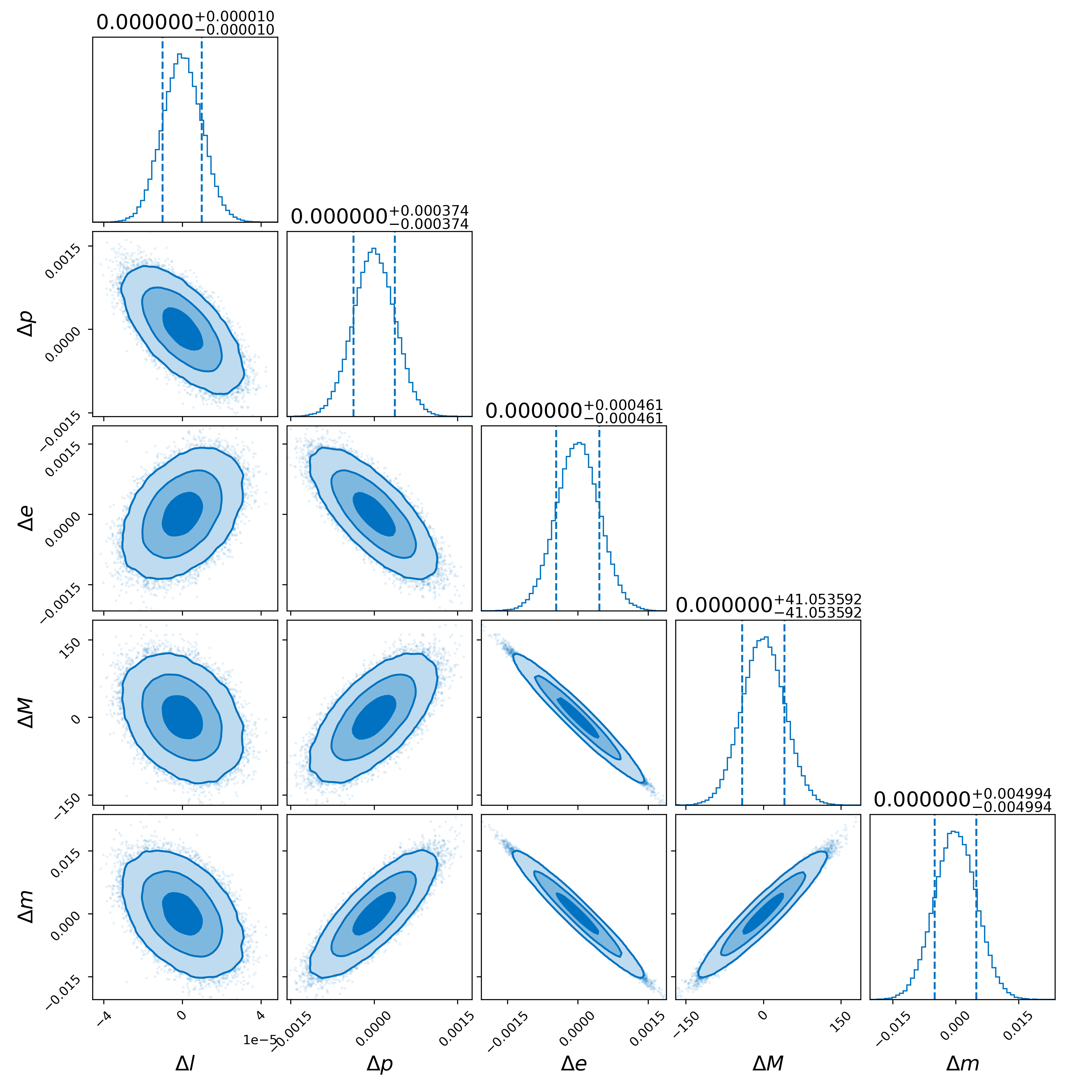}
	\end{minipage}
	\caption{\label{fig:corner} 
		The corner plot for the probability distribution of the parameters of an EMRI system. The parameters considered include the mass $(M, m)$, initial orbital eccentricity $e$, semi-latus rectum $p$, and LSB parameter $l$. The data, inferred from the one-year observation, reveals the confidence intervals for 68\%, 95\% and 99\% of the parameters, with vertical lines denoting the 1-\(\sigma\) interval for each parameter.
	}
\end{figure}


For the EMRI systems, where the inspiral occurs over a long timescale, the waveform depends sensitively on several parameters such as the mass and orbital configuration of the system. The FIM can be used to estimate the precision with which these parameters can be extracted from the gravitational wave data~\cite{vallisneri2008use}. The FIM is defined as the second derivative of the log-likelihood function with respect to the parameters of interest. Mathematically, it is expressed as
\begin{equation}
	\Gamma_{ij} = \left\langle \frac{\partial h}{\partial \theta_i} \middle| \frac{\partial h}{\partial \theta_j} \right\rangle,
\end{equation}
where \( \partial h / \partial \theta_i \) represents the partial derivative of the waveform with respect to the parameter \( \theta_i \), and the inner product is the same as that used in the definition of the overlap function. The FIM provides a local approximation to the likelihood function around the true values of the parameter, and its inverse, \( \Sigma_{ij} = (\Gamma^{-1})_{ij} \), gives an estimate of the covariance of the parameter uncertainties. The diagonal elements of the covariance matrix \( \Sigma_{ij} \) provide the expected uncertainties in each parameter
\begin{equation}
	\sigma_i = \sqrt{(\Sigma)_{ii}},
\end{equation}
where \( \sigma_i \) represents the measurement error  of the parameter \( \theta_i \). The FIM is particularly useful in the high SNR regime, where the likelihood function around the true parameters is well-approximated by a Gaussian distribution. The FIM is instrumental in setting bounds on physical parameters, such as the LSB parameter in the KR black hole model, where LISA can place the constraints on such parameters within very small fractional errors, providing insights into the nature of compact objects in the universe~\cite{vallisneri2008use,maselli2022detecting,zhang2024probing}.

Now we are in a position to compute the FIM for the constraint on the LSB parameter $l$. In our calculations, we set the SNR to 20. In Fig.~\ref{fig:corner}, we present the corner plot obtained from the FIM analysis, which illustrates the probability distributions and correlations between the intrinsic parameters $(l, M, m, p, e)$. The marginalized one-dimensional distributions along the diagonal show the spread of each parameter, while the off-diagonal panels depict the two-dimensional correlations with confidence contours corresponding to the 68\%, 95\% and 99\% probability intervals.
The results indicate that the detection error for the parameter $l$ is approximately \(\Delta l \approx 1.0 \times 10^{-5}\) at \(\mathrm{SNR} = 20\), reflecting the high sensitivity of the Fisher matrix estimation. Notably, the parameter $l$ demonstrates strong correlations with other intrinsic parameters. Specifically,  $l$ shows an inverse correlation with \(p\), \(M\) and \(m\), while it exhibits a positive correlation with \(e\). Furthermore, all parameters maintain sufficiently small uncertainties, ensuring the detectability of LSB parameter $l$.

\section{Conclusions}
\label{section:conclusion}

An EMRI system composed of a supermassive black hole and a stellar-mass compact object is expected to emit the GWs with waveforms containing $10^4$ to $10^5$ cycles within the detector sensitive frequency band, providing unique advantages for exquisite measurements of system parameters. 
The accumulation of a large number of orbital cycles can amplify slight deviations from GR, giving an unprecedented opportunity to probe the nature of the supermassive black hole. In this study, we estimate the potential of EMRIs to probe LSB effects by the KR black hole. Our results show the LSB effect appears in the leading order for the corrections of energy and angular momentum fluxes,  which is different from the loop quantum gravity corrections~\cite{Fu:2024cfk} or string theory modifications~\cite{Zhang:2024csc} considered in previous studies, where they appear at subleading orders in the energy flux. Obviously, this indicates a significant impact of the LSB effect on the orbital evolution. We notice that as the observation time increases, the LSB effects gradually alter the orbital dynamics and waveforms, which implies that the impact of the LSB effect on the EMRIs becomes increasingly significant, leading to the detectable deviations.

Specifically, we explore the impact of the KR field on the dynamics of the EMRI system. Our research suggests that the presence of KR field influences the eccentric orbital motion of the secondary object in the EMRI system, particularly the rate of change of orbital parameters like the semi-latus rectum and eccentricity. These changes manifest as detectable differences in the gravitational waveforms, with the potential for detection through  LISA. Moreover, as the LSB parameter $l$ increases, the differences in EMRI waveforms induced by the KR field become more pronounced, which makes it easier for LISA to detect the KR field. Therefore, LISA can effectively recognize the difference between Schwarzschild and KR black hole spacetimes, thereby allowing for the identification of different EMRI sources.

In addition, we estimate the detectability of the LSB effect and constrain the LSB parameter using the FIM. Based on the mismatch analysis, we demonstrate that the LSB effect becomes detectable for values of $|l|\sim 10^{-6}$ with a one-year
observation period, highlighting the sensitivity of future detectors to KR field corrections. We find that the combination of the LSB effect and the eccentricity provides richer physics in the EMRI waveforms, and the higher eccentricity makes it more easy for the detection of LSB effect. Furthermore, through the FIM, we note that the detection error for $l$ can be constrained to \(\Delta l \approx 1.0 \times 10^{-5}\) at \(\mathrm{SNR} = 20\), highlighting the potential of space-based gravitational wave detectors like LISA to rigorously test deviations from general relativity. This level of precision is crucial for constraining non-GR model parameters and testing their predictions with the observations.

\acknowledgments

The authors thank Dr. Sheng Long, Guoyang Fu and Jianjun Song for helpful discussions. This work was supported by the National Key Research and Development Program of China (Grant No. 2020YFC2201400) and National Natural Science Foundation of China (Grant Nos. 12275079, 12447156 and 12035005), and the innovative research group of Hunan Province under Grant No. 2024JJ1006.

\bibliographystyle{apsrev4-2}
\bibliography{reference}

\end{document}